\documentclass[a4paper,11pt]{article}

\usepackage[textwidth=18cm, textheight=26cm]{geometry}
\usepackage[utf8]{inputenc}
\usepackage[english]{babel}
\usepackage[T1]{fontenc}
\usepackage{graphicx}
\usepackage{lmodern}
\usepackage[hidelinks]{hyperref}
\usepackage{amsmath}
\usepackage{amssymb}
\usepackage{bm}
\usepackage{siunitx}
\usepackage{cleveref}
\usepackage[affil-it]{authblk}
\usepackage{titlesec}
\usepackage{natbib}
\usepackage{subfig}
\usepackage{booktabs}

\makeatletter
\patchcmd{\@maketitle}{\LARGE \@title}{\fontsize{14}{16.8}\selectfont\@title}{}{}
\makeatother

\titleformat{\section}{\normalfont\fontsize{10}{12}\bfseries}{\thesection}{0.5em}{}
\titleformat{\subsection}{\normalfont\fontsize{10}{12}\itshape}{\thesubsection}{0.5em}{}
\titleformat{\subsubsection}{\normalfont\fontsize{10}{12}\itshape}{\thesubsubsection}{0.5em}{}

\hypersetup{pdfauthor=author}

\begin{document}

\newcommand{\mtm}{\textit{mode-to-mode} }
\newcommand{\mts}{\textit{mode-to-shell} }
\newcommand{\sts}{\textit{shell-to-shell} }
\renewcommand{\k}{\bm{\kappa}}
\newcommand{\kP}{\bm{\kappa'}}
\newcommand{\kmkP}{\bm{\kappa} - \bm{\kappa'}}
\newcommand\Rey{\mbox{\textit{Re}}}  

\title{The influence of energy-containing scales on the distribution of spectral energy transfers}

\author[a,b]{Arthur Couteau\footnote{Corresponding author: \href{mailto:acouteau@ethz.ch}{\texttt{acouteau@ethz.ch}}}}
\author[b]{Panayotis Dimopoulos Eggenschwiler}
\author[a]{Patrick Jenny}

\affil[a]{Institute of Fluid Dynamics, ETH Zurich, CH-8092 Zürich, Switzerland}
\affil[b]{Chemical Energy Carriers and Vehicle Systems Laboratory, Empa, CH-8600 Dübendorf, Switzerland}

\maketitle
\begin{abstract}
    We present computations of individual mode-to-mode energy transfers from direct numerical simulations of homogeneous isotropic turbulence. Unlike previous approaches based on shell-filtered velocity fields, this method distinguishes between the energy exchanged by each pair of modes within a triad. We introduce a potential function based on the energy content of the modes involved and show that it predicts the distribution of intense energy transfers in the vicinity of the sampling mode considered. By performing simulations with forcing applied at intermediate wavenumbers, we demonstrate that the region of most intense transfers is determined by the spectral location of the energy-containing scales rather than by the local or nonlocal character of the triad. Direct energy exchanges with the energy-containing range are suppressed by geometric constraints from the divergence-free condition, but persist as residuals when the sampling mode is close to the energy-containing scales. The comparison with an estimator derived from EDQNM theory shows good agreement and recovers the forward, scale-local nature of energy transfer consistent with the cascade picture.
\end{abstract}

\section{Introduction}\label{sec:intro}

The energy cascade has been a cornerstone concept in turbulence theory, where energy is supplied to the system at large scales and dissipated at small scales. The classical Kolmogorov picture (K41) says that between those two ends of the spectrum, the energy is transported locally from scale to scale. As the energy travels from the large scales, where it is created to the small scales, where it is dissipated, the information contained about the large scale flow eventually gets lost and the statistics become isotropic (Kolmogorov's local isotropy hypothesis). In spectral space, the \mtm energy transfer between two modes $\k$ and $\kP$ implies a third mode, $\kmkP$, such that they form a triangle called a triad interaction. As the idea of scale locality of energy transfer is that only scales similar to a given scale contribute to energy transfer across the latter, it is expected that energy transfer occurs due to local triads, where $|\k| \approx |\kP| \approx |\kmkP|$.

An early investigation of spectral locality came from Kraichnan \citep{kraichnanInertialrangeTransferTwo1971}. The introduction of a parameter $v$, measuring the ratio of the smallest to the middle wavenumber in a triad, in the Direct Interaction Approximation \citep{kraichnanStructureIsotropicTurbulence1959} closure lead to the discovery that a significant part of the energy transfer is due to interactions in which $v < 0.5$. While this may seem contradictory, the locality of a triad, satisfied for $|\k| \approx |\kP| \approx |\kmkP|$, is not to be confused with the locality of energy transfer, which states that energy transfer should be high between neighbouring scales $|\k| \approx |\kP|$. Indeed, these early results suggested that nonlocal triads consisting of two neighbouring scales and a large scale (small wavenumber) are crucial for energy transfer.

Domaradzki and coworkers \citep{domaradzkiEnergyTransferIsotropic1988,domaradzkiLocalEnergyTransfer1990,yeungResponseIsotropicTurbulence1991,domaradzkiNonlocalTriadInteractions1992,ohkitaniTriadInteractionsForced1992} explicitly calculated band to band energy transfer functions using filtered velocity fields obtained from Direct Numerical Simulations (DNS). They showed that the most intense energy transfers occurred between neighbouring wavenumbers, mediated by a small wavenumber, thus establishing the concept of ``local energy transfer in nonlocal triad interaction''. While it is in agreement with the classical picture of the energy cascade with local transfers, this point of view clashes with another assumption of the Kolmogorov framework, that is the universality of small scales. Indeed, the statistical independence of small scales from large scale flow is questioned by the fact that the most intense triad interactions include the large scales.

Subsequent studies \citep{waleffeNatureTriadInteractions1992,zhouDegreesLocalityEnergy1993,zhouInteractingScalesEnergy1993}, while confirming the predominance of nonlocal triads in the energy flux, argued that the \mtm energy transfers are not the appropriate quantities to determine the locality of energy transfer. Instead, integrated quantities like the \sts energy transfer functions should be investigated. In this aggregate function, the influence of the nonlocal interactions producing intense local energy transfers diminishes. Indeed, there is a fixed number of nonlocal interactions compared to the total number of interactions, and further cancellations due to sign definite regions are expected. \citet{eyinkEnergyDissipationViscosity1994,eyinkLocalEnergyFlux1995} introduced another definition of local flux to show that local interactions do dominate in the limit of high Reynolds number, in order to justify unversality of small scale statistics. Later, they extended the previous work for non-sharp filters and obtained bounds on the contribution of nonlocal interactions to the local energy flux \citep{eyinkLocalityTurbulentCascades2005,eyinkLocalnessEnergyCascade2009,aluieLocalnessEnergyCascade2009}.

Following works \citep{alexakisImprintLargeScaleFlows2005,mininniLargescaleFlowEffects2006,domaradzkiAnalysisEnergyTransfer2007,domaradzkiComparisonSpectralSharp2007,domaradzkiLocalityPropertiesEnergy2009} continue to support the ``local energy transfer in nonlocal triad interaction'' perspective via higher Reynolds number simulations, while showing a weakening, but non-vanishing, of these interactions as the $\Rey$ is growing \citep{mininniNonlocalInteractionsHydrodynamic2008}. With the use again of statistics based on spectral band-truncated velocity fields, they showed the coexistence of nonlocal interactions and local energy cascade. Furthermore, they highlighted that the extent of locality, that is the number of bands involved in intense local energy transfers, is dependent on the numerical forcing conditions.

Furthermore, while not directly looking at the locality question, recent studies \citep{kangAlignmentsTriadPhases2021,protasAlignmentsTriadPhases2024} investigate the ``path'' of energy across spectral modes and show that phase correlation in phase space is also local, suggesting that energy is transferred locally between neighbouring modes.

While the locality of energy transfer is established in all works mentioned above, there is controversy about which triad interactions are responsible for it, the main question being ``are the large scales involved in energy transfer at any other scale?''. We would like to put this question in the general context of the statistical independence of small scales, as assumed by Kolmogorov theory for large enough $\Rey$. In this regard, there are still ongoing discussions about deviations from the expected behaviour of universal small scale statistics. For example, a slower return to isotropy or even a persistence of anisotropy at small scales from large scale forcing was observed numerically \citep{biferaleAnisotropicHomogeneousTurbulence2001,biferaleIsotropyVsAnisotropy2001} and experimentally \citep{carterScaletoscaleAnisotropyHomogeneous2017,carterSmallscaleStructureEnergy2018}. These are usually attributed to moderate Reynolds number effects, see \citet{iyerReynoldsNumberScaling2017} for a discussion on the topic.

In the present work, we compute individual mode-to-mode energy transfers directly from DNS, which allows us to distinguish between the energy exchanged by each pair of modes within a triad. We introduce a potential function based on the energy content of the modes involved and show that it predicts the distribution of intense energy transfers. By performing simulations with forcing applied at intermediate wavenumbers, we demonstrate that the most intense transfers are determined by the spectral location of the energy-containing range rather than by the local or nonlocal character of the interaction. The methodology and definitions are presented in Sec. \ref{sec:method}, followed by results for classical large-scale forcing in Sec. \ref{sec:modeToModeEnergyTransfer} and for shifted forcing configurations in Sec. \ref{sec:ExtentOfLocality}.

\section{Methodology}\label{sec:method}
\subsection{Incompressible flow in spectral space}

The Navier-Stokes equations for an incompressible flow are given by
\begin{align}
     & \partial_{t} u_{j} + u_{l} \frac{\partial u_{j}}{\partial x_{l}} = - \frac{\partial p}{\partial x_{j}} + \nu \frac{\partial^{2} u_{j}}{\partial x_{l} \partial x_{l}}\quad \mathrm{and} \\
     & \frac{\partial u_{j}}{\partial x_{j}} = 0.
\end{align}
Using the Fourier series representation of the velocity field $u_{j}(\bm{x})$ as
\begin{equation}
    u_{j}(\bm{x}) = \sum_{\bm{\kappa}} \hat{u}_{j}(\bm{\kappa}) e^{i \bm{\kappa} \cdot \bm{x}},
\end{equation}
the Navier-Stokes equations are transformed to spectral space, reading
\begin{align}
     & (\partial_{t} + \nu \kappa^{2}) \hat{u}_{j}(\bm{\kappa}) = - P_{jm}(\bm{\kappa}) \hat{C}_{m}(\bm{\kappa})\quad \mathrm{and} \\
     & \kappa_{j} \hat{u}_{j}(\bm{\kappa}) = 0,\label{eq:DivFree}
\end{align}
where the pressure projection tensor $P_{jm}(\bm{\kappa})$ is given by
\begin{equation}
    P_{jm}(\bm{\kappa}) = \delta_{jm} - \frac{\kappa_{j} \kappa_{m}}{\kappa^{2}},
\end{equation}
and the convection term is expressed as a convolution product $\hat{C}_{m}(\bm{\kappa}) = \sum_{\bm{\kappa'}} \hat{c}_{m}(\bm{\kappa}, \bm{\kappa'})$, with the single contributions reading
\begin{equation}
    \hat{c}_{m}(\bm{\kappa}, \bm{\kappa'}) = i \kappa_{l} \hat{u}_{m}(\bm{\kappa'}) \hat{u}_{l}(\bm{\kappa} - \bm{\kappa'}).
\end{equation}
The kinetic energy of mode $\bm{\kappa}$ is defined as $\hat{E}(\bm{\kappa}) = 1/2 \ \hat{u}_{j}(\bm{\kappa}) \hat{u}_{j}^{*}(\bm{\kappa})$, where $^{*}$ denotes the complex conjugate, and its evolution follows directly from the Navier-Stokes equations as
\begin{equation}
    (\partial_{t} + \nu \kappa^{2}) \hat{E}(\bm{\kappa}) = - \sum_{\bm{\kappa'}} \hat{T}(\bm{\kappa}, \bm{\kappa'}),
    \label{eq:modalEnergyConservation}
\end{equation}
with the \mtm energy transfer rate
\begin{equation}
    \hat{T}(\bm{\kappa}, \bm{\kappa'}) = \Re\left\{ P_{jm}(\bm{\kappa}) \hat{c}_{m}(\bm{\kappa}, \bm{\kappa'}) \hat{u}_{j}^{*}(\bm{\kappa}) \right\} = \Re\left\{ i \kappa_{l} \hat{u}_{l}(\bm{\kappa} - \bm{\kappa'}) \hat{u}_{j}(\bm{\kappa'})  \hat{u}_{j}^{*}(\bm{\kappa}) \right\}. \label{eq:ModeToModeEnergyTransfer}
\end{equation}
This quantity represents the energy exchange rate between modes $\bm{\kappa}$ and $\bm{\kappa'}$. A triad interaction consists of three \mtm energy transfers, one between each pair of modes involved. By convention, a negative $\hat{T}(\bm{\kappa}, \bm{\kappa'})$ means energy transfer from $\bm{\kappa'}$ to $\bm{\kappa}$ (gain) and a positive $\hat{T}(\bm{\kappa}, \bm{\kappa'})$ means energy transfer from $\bm{\kappa}$ to $\bm{\kappa'}$ (loss).

The quantity of reference to discuss the locality of energy transfer is the \sts transfer function. A shell $K$ is defined as a set of modes defined by a radius $k$ and an increment $\Delta k$ as $K = \left\{ \bm{\kappa'} \mid  k \le |\bm{\kappa'}| < k + \Delta k \right\}$. The \sts transfer function, that is, the total amount of energy transfer between shell $K$ and $K'$, is given by integration of $\hat{T}$ as
\begin{equation}
    \hat{S}(K, K') = \sum_{\bm{\kappa} \in K} \sum_{\bm{\kappa'} \in K'} \hat{T}(\bm{\kappa}, \bm{\kappa'}).
    \label{eq:shellToShell}
\end{equation}
This requires knowledge of the whole $(\bm{\kappa}, \bm{\kappa'})$ space, which is 6-dimensional and therefore not directly computable. Instead, we will present computations of $\hat{T}$ for given sampling wavenumbers $\bm{\kappa}^{(s)}$, for all $\bm{\kappa'}$. We argue here that thanks to isotropy the \mtm energy transfer $\hat{T}(\bm{\kappa}^{(s)}, \bm{\kappa'})$ is similar up to a rotation for two different sampling points if they have the same norm, as they imply summation over a shell. Therefore, we will compute the \mts energy transfer as
\begin{equation}
    \hat{S}(\bm{\kappa}, K') = \sum_{\bm{\kappa'} \in K'} \hat{T}(\bm{\kappa}, \bm{\kappa'}),
    \label{eq:modeToShell}
\end{equation}
which on average is representative of $\hat{S}(K, K')$ (the expected value is recovered upon multiplication with the number of modes in shell $K$).

\subsection{Locality of energy transfers}
\begin{figure}[t]
    \centering
    \subfloat[{}\label{subfig:localInteraction}]{%
        \includegraphics[]{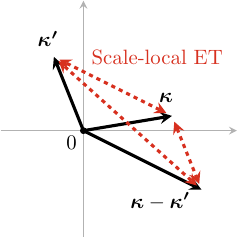}
    }
    \hspace{2cm}
    \subfloat[{}\label{subfig:nonlocalInteraction}]{%
        \includegraphics[]{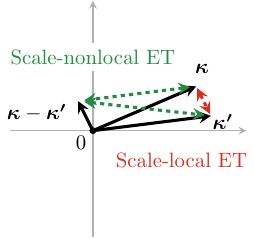}
    }
    \caption{Scale-local energy transfer (ET) within a local interaction (a) and nonlocal interaction (b).}\label{fig:localVSnonlocal}
\end{figure}

\begin{figure}[t]
    \centering
    \subfloat[{}\label{subfig:ModelSpectrumLWFEnergySpectrum}]{%
        \includegraphics[]{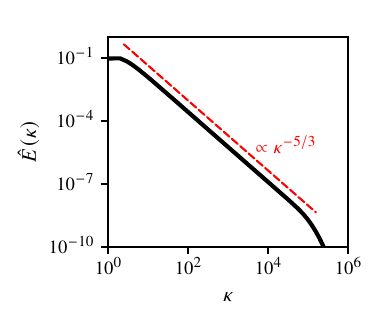}
    }
    \hspace{2cm}
    \subfloat[{}\label{subfig:ModelSpectrumLWFPotentialLinear}]{%
        \includegraphics[]{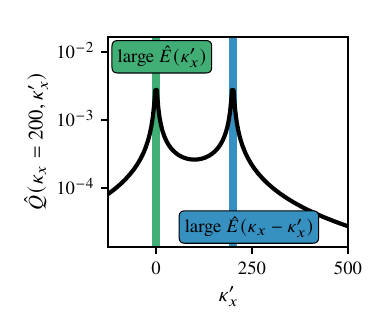}
    }
    \\
    \subfloat[{}\label{subfig:ModelSpectrumLWFPotentialLog}]{%
        \includegraphics[]{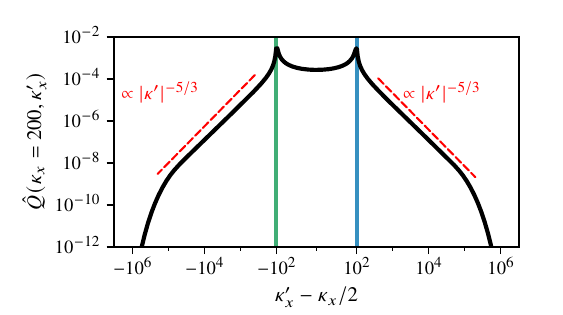}
    }
    \caption{Model kinetic spectrum with energy at largest scales (a), resulting \mtm potential $\hat{Q}$ (b) and in log scale to show the decay of potential (c). Note that (c) is centered around $\kappa_{x}/2$ to highlight the symmetry of $\hat{Q}$.}\label{fig:ModelSpectrumLWFScaling}
\end{figure}

Historically, two notions of locality have emerged. The first one, which we will refer to as ``Kraichnan locality'', derives from the assumption of universality of small scales and energy cascade. To satisfy this hypothesis, the scales involved in the energy transfer across a given scale have to be similar ($|\k| \approx |\kP| \approx |\kmkP|$), as shown in Fig. \ref{subfig:localInteraction}.

The second notion comes from the first statistics gathered from numerical simulations dedicated to the locality topic. Without access to the triad interaction statistics, \citet{domaradzkiLocalEnergyTransfer1990} gathered information using transfer functions based on the quantity $T(k|p,q)$, describing the effect on scales of wavenumber $k$ of all triad interactions involving wavenumbers of size $p$ and $q$. They showed that most of the energy transfer was due to triads with a leg in the large scales, see Fig. \ref{subfig:nonlocalInteraction}, hence introducing the concept of ``local energy transfer in nonlocal interactions''. These results, while acknowledging the locality of energy transfer, clash with the classical theory as the most active interactions involve a large scale.

While the two proposals agree on the locality of the energy cascade, there is controversy about which interactions are responsible for it. Theoretical studies predict an upper bound on the contributions of nonlocal interactions \citep{eyinkLocalEnergyFlux1995,eyinkLocalityTurbulentCascades2005,aluieLocalnessEnergyCascade2009}, while numerical results point that nonlocal interactions with a leg in the large scales are determinant for energy transfer. \citet{mininniNonlocalInteractionsHydrodynamic2008} numerically showed a weakening but non-vanishing influence of nonlocal interactions against increasing $\Rey$.

In the present work, we present results obtained from DNS with moderate Reynolds number, which confirm the dominance of local energy transfer in nonlocal interactions at low Reynolds number numerically observed previously. We propose that these kinds of interactions will always dominate in intensity (but not in number) and introduce for that purpose the potential of the \mtm energy transfer, that is the absolute maximum energy transfer between modes $\bm{\kappa}$ and $\bm{\kappa'}$. We derive this expression based on dimensional arguments from the expression for the \mtm energy transfer $\hat{T}(\bm{\kappa}, \bm{\kappa'})$ (Eq. \ref{eq:ModeToModeEnergyTransfer}), relating the velocity modes involved to their energy content
\begin{equation}
    \hat{Q}(\bm{\kappa}, \bm{\kappa'}) = |\bm{\kappa}| \hat{E}(\bm{\kappa})^{1/2} \hat{E}(\bm{\kappa'})^{1/2} \hat{E}(\bm{\kappa} - \bm{\kappa'})^{1/2}.
    \label{eq:ModeToModeEnergyTransferPotential}
\end{equation}
Let us note that $\hat{T}(\bm{\kappa}, \bm{\kappa'})$ (and corresponding $\hat{Q}$) represent the energy transfer rate between $\bm{\kappa}$ and $\bm{\kappa'}$ and that the mode $\bm{\kappa} - \bm{\kappa'}$ does not directly exchange energy with $\bm{\kappa}$ within this \mtm transfer. We therefore call $\bm{\kappa}$ the primary (or sampling) mode, $\bm{\kappa'}$ the \textit{reacting} mode, and $\bm{\kappa} - \bm{\kappa'}$ the \textit{catalyst} mode.

Of course, we do not mean that the energy transfer $\hat{T}(\bm{\kappa}, \bm{\kappa} - \bm{\kappa'})$ is identically zero, but that it should be considered as a separate energy transfer for which $\bm{\kappa} - \bm{\kappa'}$ is the \textit{reacting} mode (exchanging energy) and $\bm{\kappa'}$ the \textit{catalyst} mode (not exchanging energy). While both \mtm transfers involve the same triangle of modes and are usually considered as a single triad interaction in the literature, we make the distinction based on which modes can gain/lose energy and which cannot.

To investigate $\hat{Q}$, let us assume a classical scenario, with the ECR confined at the largest scales. For that purpose we use the model spectrum from \citet{pope2000turbulent}, determined by the energy containing scale $L = L_{0} = 2\pi$, the viscosity $\nu = 10^{-7}$ and $\Rey_{\lambda} = 10^{5}$, as drawn in Fig. \ref{subfig:ModelSpectrumLWFEnergySpectrum}.

With a fixed primary mode $\bm{\kappa}$, the potential depends on the product of the square root of the energy of both the \textit{reacting} and \textit{catalyst} modes. This function is symmetric and exhibits two peaks, as highlighted in Fig. \ref{subfig:ModelSpectrumLWFPotentialLinear}. They correspond to regions of high potential $\hat{Q}$, where either the \textit{reacting} (green) or the \textit{catalyst} (blue) modes have high energy content (i.e., are in the ECR). Note that the potential is determined by the spectrum and its scaling with $|\bm{\kappa}|$, and that a scaling exponent other than $-5/3$ may result in a different distribution of regions with high potential.

The first region of interest (blue) highlights the choice of the \textit{catalyst} terminology. Indeed, in this region the \textit{catalyst} mode has a high energy content and will enhance the energy transfer between $\bm{\kappa}$ and $\bm{\kappa'}$. As the energy is at the largest scales, the dominant interactions will be for $\bm{\kappa} - \bm{\kappa'} \approx 0$, so for $\bm{\kappa} \approx \bm{\kappa'}$. Therefore this region corresponds to the nonlocal interactions showcasing intense local energy transfer, as reported in the literature \citep{domaradzkiLocalEnergyTransfer1990}. Assuming for now that $\hat{Q}$ is a good estimator for $\hat{T}$ in this region, we can reformulate the locality statement: the dominant local energy transfer observed is not due to the nonlocal nature of the interactions, but rather because the nonlocal interactions involve a \textit{catalyst} mode in the ECR. This shifts the understanding of the locality of the intense energy transfers from the locality/non-locality of a triad to the energy content of its modes, as hinted in \citet{alexakisImprintLargeScaleFlows2005} and \citet{mininniLargescaleFlowEffects2006} in their discussion about the dependence of the energy transfer function to the forcing range. This proposal will also let us discuss the cases where the ECR is not located at the largest scales in Sec. \ref{sec:ExtentOfLocality}.

The other region of high energy transfer potential (green) is the case where the \textit{reacting} mode has a high energy content. Due to the symmetry of the definition of $\hat{Q}$, it corresponds to nonlocal energy transfer in nonlocal interactions. In other words, a mode $\bm{\kappa}$ has the potential to directly exchange energy with modes in the ECR, wherever it is. While it has been reported that energy transfer functions show a slight positive exchange with the large scales \citep{domaradzkiLocalEnergyTransfer1990}, these kinds of energy transfer have never been observed, and would go against the established picture of the scale-localness of the energy cascade. We will discuss dampening of these \mtm transfers in Sec. \ref{sec:LWFCatalyticVSReacting} and residual nonlocal energy transfers in Sec. \ref{sec:ResidualNonlocalEnergyTransfers}.

Assuming that both modes are in the inertial range, such that $\hat{E}(\bm{\kappa}) \propto |\bm{\kappa}|^{-5/3}$, the scaling of the potential goes like
\begin{equation}
    \hat{Q}(\bm{\kappa}, \bm{\kappa'}) \propto |\bm{\kappa'}|^{-5/6} |\bm{\kappa} - \bm{\kappa'}|^{-5/6}.
    \label{eq:scalingPotential}
\end{equation}
Furthermore, if $|\bm{\kappa'}| \gg |\bm{\kappa}|$, then $|\bm{\kappa} - \bm{\kappa'}| \approx |\bm{\kappa'}|$ and the scaling reduces to
\begin{equation}
    \hat{Q}(\bm{\kappa}, \bm{\kappa'}) \propto |\bm{\kappa'}|^{-5/3}.
    \label{eq:scalingPotentialLimit}
\end{equation}
This scaling behaviour is highlighted in Fig. \ref{subfig:ModelSpectrumLWFPotentialLog}, between the regions of high potential and the regions where either the \textit{reacting} or \textit{catalyst} mode is in the dissipation range.

\subsection{Transfer functions from the literature}
\label{sec:ComparisonLiterature}

As mentioned in the introduction, the numerical research on the topic of locality \citep{domaradzkiLocalEnergyTransfer1990,alexakisImprintLargeScaleFlows2005} used a second-order quantity to study locality. To obtain the latter, they calculated the third-order correlator $T_{3}$, defined as
\begin{equation}
    T_{3}(K, P, Q) = \int \bm{u}_{K} \cdot (\bm{u}_{P} \cdot \nabla) \bm{u}_{Q} d\bm{x}^{3},
\end{equation}
where $\bm{u}_{K}$ stands for a velocity field filtered to retain only the modes in shell $K$. It represents ``the rate of energy transfer from energy in a shell $Q$ to energy in shell $K$ due to interaction with the velocity field in shell $P$'' \citep{mininniLargescaleFlowEffects2006}. The major difference between $T_{3}$ and the \mtm energy transfer $\hat{T}$ (Eq. \ref{eq:ModeToModeEnergyTransfer}) is that ``this term [$T_{3}$] does not give information about the energy the shell $P$ receives or gives to shells $K$ and $Q$'' \citep{mininniLargescaleFlowEffects2006}. Therefore, it is not able to differentiate energy transfer between $P$ and $K$ and between $Q$ and $K$, as it is the indistinguishable sum of all interactions involving the retained modes. On the contrary, $\hat{T}(\bm{\kappa}, \bm{\kappa'})$ is by definition the transfer between $\bm{\kappa}$ and $\bm{\kappa'}$, such that $\hat{T}(\bm{\kappa}, \bm{\kappa'})$ and $\hat{T}(\bm{\kappa}, \bm{\kappa} - \bm{\kappa'})$, while involving the same triangle of modes, can be studied separately. Hence our distinction between \textit{reacting} and \textit{catalyst} modes, which could not be done with the quantities used in the literature.

\begin{figure}
    \centering
    \includegraphics[]{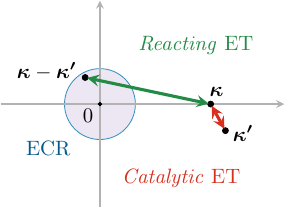}
    \caption{Distinction within a triad interaction with a mode in the energy containing range of the \textit{catalytic} and \textit{reacting} \mtm energy transfers (ET)}\label{fig:nonlocalSeparated}
\end{figure}

This distinction is particularly important in the case of triad interaction such that one mode is in the energy containing range (ECR), as shown in Fig. \ref{fig:nonlocalSeparated}. Within this triad interaction, we identify the \textit{reacting} \mtm transfer as the energy transfer between the primary mode and the \textit{reacting} mode in the ECR, mediated by the \textit{catalytic} mode in the vicinity of the primary mode, and the \textit{catalytic} \mtm transfer as the energy transfer between the primary mode and the \textit{reacting} mode in the vicinity of the primary mode, mediated by the \textit{catalytic} mode in the ECR. By convention, the primary mode is called $\k$, the mode in the vicinity $\kP$ and the mode in the ECR $\kmkP$. Both transfers have the same potential by construction, $\hat{Q}(\k, \kP) = \hat{Q}(\k, \kmkP)$. However, the energy transfer rates are different, $\hat{T}(\k, \kP) \ne \hat{T}(\k, \kmkP)$.

The limitation of $T_{3}$ is overcome by integration over all $P$ shells, denoted in the literature as $T_{2}(K, Q) = \sum_{P} T_{3}(K, P, Q)$ (similar then to the second-order correlator $T(k|p,q)$ from \citep{domaradzkiLocalEnergyTransfer1990}). The summation removes the ambiguity of the energy transfer to/from $P$ shells, and is similar to the \mts estimate $\hat{S}(\bm{\kappa}, K')$ (Eq. \ref{eq:modeToShell}).

\subsection{Comparison with theoretical predictions}
\label{sec:comparisonTheoreticalPredictions}

\begin{figure}[t]
    \centering
    \subfloat[{$\hat{Q}(\bm{\kappa} = [200,0], \bm{\kappa'})$}\label{subfig:ModelSpectrumLWFPotentialSlice}]{%
        \includegraphics[]{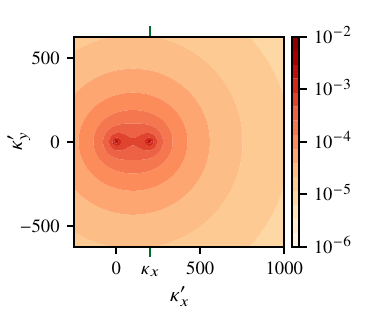}
    }
    \hspace{2cm}
    \subfloat[{$\hat{H}(\bm{\kappa} = [200,0], \bm{\kappa'})$}\label{subfig:ModelSpectrumLWFEDQNMSlice}]{%
        \includegraphics[]{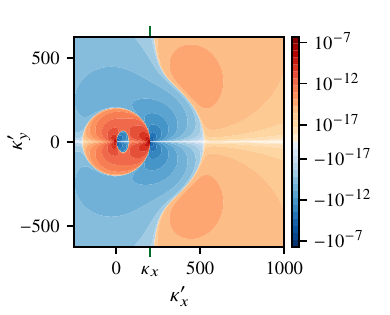}
    }
    \caption{(a) Potential $\hat{Q}$ and (b) EDQNM estimator $\hat{H}$ computed from the model spectrum for $\bm{\kappa} = [200,0]$.}\label{fig:ModelSpectrumLWFSlices}
\end{figure}

A powerful theoretical tool to quantify the energy transfers is provided by the analytical theories of turbulence introduced in the 70s, like the Direct Interaction Approximation \citep{kraichnanStructureIsotropicTurbulence1959} or Eddy-Damped Quasi Normal Markovian (EDQNM) \citep{orszagAnalyticalTheoriesTurbulence1970} approaches. We chose an estimator $\hat{H}(\bm{\kappa}, \bm{\kappa'})$ of the \mtm energy transfer derived from the EDQNM theory as presented in \citet{lesieurTurbulenceFluids2008}, which we summarize below
\begin{equation}
    \label{eq:EDQNMH}
    \hat{H}(\bm{\kappa}, \bm{\kappa'}) = \theta(\bm{\kappa}, \bm{\kappa'}) b(\bm{\kappa}, \bm{\kappa'}) \frac{|\bm{\kappa}|}{|\bm{\kappa'}||\bm{\kappa} - \bm{\kappa'}|} \hat{f}(\bm{\kappa}, \bm{\kappa'}),
\end{equation}
with the function $\hat{f}(\bm{\kappa}, \bm{\kappa'})$ as
\begin{equation}
    \hat{f}(\bm{\kappa}, \bm{\kappa'}) = \hat{E}(\bm{\kappa} - \bm{\kappa'}) \left( |\bm{\kappa}|^2 \hat{E}(\bm{\kappa'}) - |\bm{\kappa'}|^2 \hat{E}(\bm{\kappa}) \right).
    \label{eq:EDQNMf}
\end{equation}
The factor $b(\bm{\kappa}, \bm{\kappa'})$ depends on the shape of the triad and is given by
\begin{equation}
    b(\bm{\kappa}, \bm{\kappa'}) = \frac{|\bm{\kappa'}|}{|\bm{\kappa}|} \left( xy + z^{3} \right),
    \label{eq:EDQNMb}
\end{equation}
where $x$, $y$ and $z$ are the cosines of the angles opposite to $\bm{\kappa}$, $\bm{\kappa'}$ and $\bm{\kappa} - \bm{\kappa'}$, respectively. The triad interaction time $\theta(\bm{\kappa}, \bm{\kappa'})$ is simplified for numerical procedure following \citep{domaradzkiLocalEnergyTransfer1990}, resulting in the function
\begin{equation}
    \theta(\bm{\kappa}, \bm{\kappa'}) = \left[ \eta(\bm{\kappa}) + \eta(\bm{\kappa'}) + \eta(\bm{\kappa} - \bm{\kappa'}) \right]^{-1},
    \label{eq:EDQNMtheta}
\end{equation}
with the function
\begin{equation}
    \eta(\bm{\kappa}) = \nu |\bm{\kappa}|^2 + 0.31 \left[ |\bm{\kappa}|^3 \hat{E}(\bm{\kappa}) \right]^{1/2},
\end{equation}
as proposed by \citet{orszagAnalyticalTheoriesTurbulence1970}. Figure \ref{fig:ModelSpectrumLWFSlices} presents slices of the potential $\hat{Q}$ and the absolute value of the EDQNM estimator $\hat{H}$ computed from the model spectrum introduced in the previous section, see Fig. \ref{subfig:ModelSpectrumLWFEnergySpectrum}. These functions both possess a cylindrical symmetry around the line in the direction of the primary mode $\bm{\kappa}$, here the $\kappa'_{x}$ axis.

Let us note that while the EDQNM estimator $\hat{H}$ is not symmetric in its arguments, it exhibits a similar pattern to that of the potential. That is, there are mainly two regions where $|\hat{H}|$ is large, locally around the primary mode as it can be expected from the local energy transfer from the nonlocal interactions, but also in the ECR, close to the origin. Now, while the potential predicts similar levels of local and nonlocal  energy transfers, the latter is much weaker for $\hat{H}$ (here approximately five orders of magnitude smaller), in agreement with the theory. However, it is interesting to note that it is still larger than the rest of the domain.

Another striking difference is that $|\hat{H}|$ is much smaller everywhere in the domain than the potential. This is to be expected, as the potential assumes that every quantity in the \mtm energy transfer $\hat{T}$ is aligned and their phases are such that the energy transfer is optimal, which cannot be the case. Furthermore, $|\hat{H}|$ decays much faster than the potential as $\bm{\kappa'}$ moves away from the primary mode.

\begin{figure}[t]
    \centering
    \subfloat[{$\hat{f}(\bm{\kappa} = [200,0], \bm{\kappa'})$}\label{subfig:ModelSpectrumLWFEDQNMfSlice}]{%
        \includegraphics[]{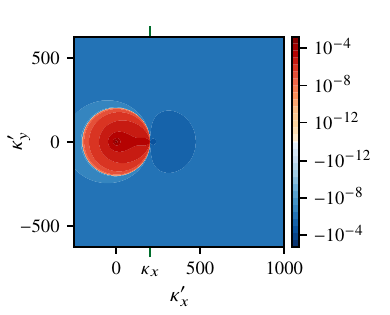}
    }
    \hspace{2cm}
    \subfloat[{$b(\bm{\kappa} = [200,0], \bm{\kappa'})$}\label{subfig:ModelSpectrumLWFEDQNMbSlice}]{%
        \includegraphics[]{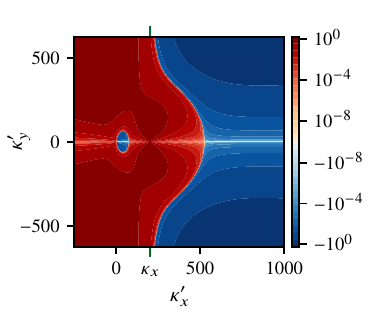}
    }
    \caption{Function $\hat{f}$ (a) and factor $b$ (b) from EDQNM estimator computed from the model spectrum for $\bm{\kappa} = [200,0]$.}\label{fig:ModelSpectrumLWFEDQNMSlices}
\end{figure}

Finally, the EDQNM estimator shows specular, i.e., angle-dependent, behaviour around the origin and the primary mode. This is due to the zeros of the function $\hat{f}$ (Eq. \ref{eq:EDQNMf}) and the factor $b$ (Eq. \ref{eq:EDQNMb}), as shown in Fig. \ref{fig:ModelSpectrumLWFEDQNMSlices}. The physical relevance of these functions will be discussed in Secs. \ref{sec:LWFDistributionOfT} and \ref{sec:IWFDistributionOfT} by comparing $\hat{H}$ to the values of $\hat{T}$ obtained from DNS.

\subsection{Numerical procedure}
\label{sec:numericalProcedure}

\begin{table}[t]
    \centering
    \begin{tabular}{@{}lccccc@{}}
        Case  & $N$ & $\nu$            & Forcing range                 & $\Rey_{\lambda}$ \\
        \midrule
        LWF   & 512 & $1.5\times10^{-3}$ & $\left] 0, 2\sqrt{2} \right]$ & $160 \sim 170$   \\
        \midrule
        IWF10 & 512 & $3\times10^{-3}$ & $\left] 9, 11 \right]$       & $\sim 60$        \\
        \midrule
        IWF20 & 512 & $4\times10^{-3}$ & $\left] 19, 21 \right]$       & $\sim 30$        \\
        \bottomrule
    \end{tabular}
    \caption{Simulation parameters}\label{tab:SimulationParameters}
\end{table}

In order to validate our hypotheses, we performed  three Direct Numerical Simulations (DNS) with a resolution of $512^{3}$, using a pseudo-spectral code. The three cases differ by the spectral forcing range, where the first one corresponds to a Low Wavenumber Forcing (LWF) configuration, with the forcing applied at the largest scales in the domain, while the two others consider forcing ranges shifted to intermediate wavenumbers (Intermediate WF). As a result of the shifted forcing range, the ratio of the integral scale to the Kolmogorov length scale is significantly reduced, resulting in a lower $\Rey_{\lambda}$. The simulation parameters are summarized in Tab. \ref{tab:SimulationParameters}.

For each case, several independent simulations were time and ensemble averaged once statistical stationarity has been reached. This results in the averaging of up to 12000 time frames gathered over a total of 800 eddy turnover times for the IWF20 case, while the LWF and IWF10 cases converged with less statistics.

The \mtm energy transfer $\hat{T}$, the potential $\hat{Q}$ and the EDQNM estimator $\hat{H}$ were computed directly from Eqs. \ref{eq:ModeToModeEnergyTransfer}, \ref{eq:ModeToModeEnergyTransferPotential} and \ref{eq:EDQNMH}, respectively, for a set of sample points chosen along the $x$ axis, from $\kappa^{(s)}_{x} = 10$ to $\kappa^{(s)}_{x} = 80$. The algorithm used to efficiently perform this computation is presented in the appendix \ref{app:EnergyTransferComputation}. The obtained energy transfer fields were validated against the energy transfers computed from the pseudo-spectral method, as the sum of all contributions has to be equal to the total energy transfer, i.e., $\sum_{\kP} \hat{T}(\k, \kP) = \hat{C}_{m}(\k) \hat{u}_{m}^{\ast}(\k)$.

\section{Spectral energy transfers in homogeneous isotropic turbulence}
\label{sec:modeToModeEnergyTransfer}
\begin{figure}[t]
    \centering
    \includegraphics[]{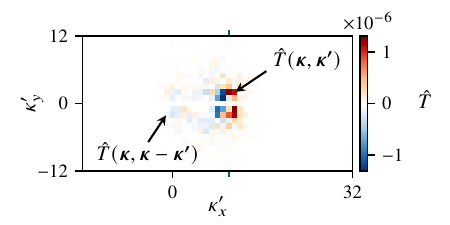}
    \caption{Distribution of energy transfers $\hat{T}(\bm{\kappa}, \bm{\kappa'})$ for the LWF case ($\kappa_{f} \sim 1$), for $\bm{\kappa} = [10,0]$. The green dashes indicate the position of the sampling point. For a given $\bm{\kappa'}$, the \mtm transfers corresponding to $\hat{T}(\bm{\kappa}, \bm{\kappa'})$ and $\hat{T}(\bm{\kappa}, \bm{\kappa} - \bm{\kappa'})$ are read as indicated.}\label{fig:LWF_M2M_Explanation}
\end{figure}

In this section, we present the results obtained from the DNS simulations in the classical case of LWF (first row of Tab. \ref{tab:SimulationParameters}). Figure \ref{fig:LWF_M2M_Explanation} presents an example of the distribution for the sampling point $\bm{\kappa} = [10,0]$. Each point represents the value of the \mtm energy transfer $\hat{T}(\bm{\kappa}, \bm{\kappa'})$ between the sampling mode and the mode given by the coordinate of the point (\textit{reacting} mode). For a chosen mode $\kP$, the \mtm energy transfer to the other mode of the triad $\kmkP$ is read directly from the same figure, as indicated. The remainder of the article is dedicated to presenting and discussing the distribution of $\hat{T}$ and the dominance of energy transfers in the vicinity of the sampling mode. All subsequent figures are zoomed in such a way that all cases presented in Tab. \ref{tab:SimulationParameters} can be directly compared.

\subsection{Distribution of spectral energy transfers}
\label{sec:LWFDistributionOfT}
In order to discuss the distribution of \mtm energy transfers, let us define the High Energy Transfer (HET) kernel as the set of modes with high energy transfer intensities.

\begin{figure}[t]
    \centering
    \subfloat[{$\bm{\kappa} = [20,0]$}\label{subfig:LWF_M2M_T_XY_20}]{%
        \includegraphics[]{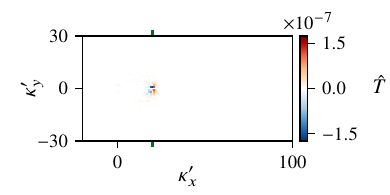}
    }
    \hspace{2cm}
    \subfloat[{$\bm{\kappa} = [40,0]$}\label{subfig:LWF_M2M_T_XY_40}]{%
        \includegraphics[]{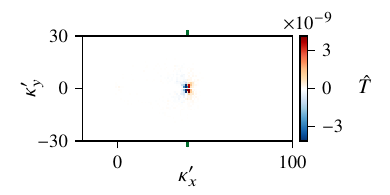}
    }
    \\
    \subfloat[{$\bm{\kappa} = [60,0]$}\label{subfig:LWF_M2M_T_XY_60}]{%
        \includegraphics[]{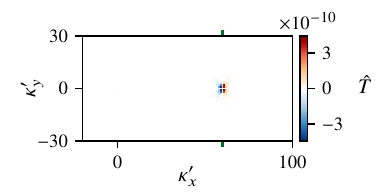}
    }
    \hspace{2cm}
    \subfloat[{$\bm{\kappa} = [80,0]$}\label{subfig:LWF_M2M_T_XY_80}]{%
        \includegraphics[]{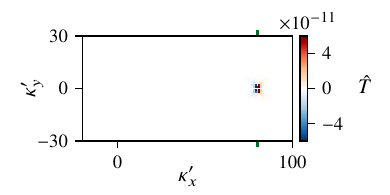}
    }
    \caption{Distribution of energy transfers $\hat{T}(\bm{\kappa}, \bm{\kappa'})$ for the LWF case ($\kappa_{f} \sim 1$). The green dashes indicate the position of the sampling point.}\label{fig:LWF_M2M}
\end{figure}

Figure \ref{fig:LWF_M2M} shows slices in the $\kappa'_{x}, \kappa'_{y}$ plane of the distribution of the \mtm energy transfer $\hat{T}(\bm{\kappa}, \bm{\kappa'})$. We find that for every sampling point, the transfers are dominant in intensity in the vicinity of the latter, i.e., the HET kernel is composed of the neighbours of the sampling point, corresponding to a sphere in three dimensions. These transfers correspond to the ``scale-local energy transfer in nonlocal interactions'', as illustrated in Fig. \ref{subfig:nonlocalInteraction}, where the sampling and \textit{reacting} modes are neighbours and the \textit{catalyst} mode is near the origin.

An interesting feature is that the HET kernel is independent of the sampling point wavenumber. Indeed, the latter is composed of approximately the same number of modes forming a spherical region (in three dimensions) around $\bm{\kappa}$. These figures also demonstrate that the energy transfer is scale local and forward on average. The modes with a smaller wavenumber than $\bm{\kappa}$ exhibit a negative $\hat{T}$ (energy gain) while the modes with larger wavenumber show a postive $\hat{T}$ (energy loss). Notably, the energy transfer due to convection is similar whether the sampling point is in the inertial or the dissipation range ($\kappa_{\eta} \approx 42$).

\begin{figure}[t]
    \centering
    \includegraphics[]{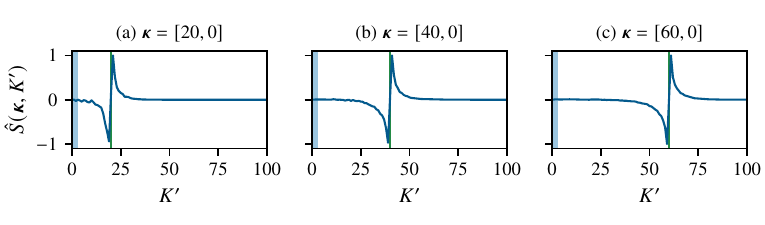}
    \caption{\textit{Shell-to-shell} transfer functions $\hat{S}(\kappa, K')$ for the LWF case ($\kappa_{f} \sim 1$). The blue shaded area correspond to the ECR and the vertical green line to the sampling point.}\label{fig:LWF_M2S}
\end{figure}

These features are made clearer when we compare the aggregated \mts transfer function $\hat{S}$ for different sampling points, shown in Fig. \ref{fig:LWF_M2S}. The number of shells relevant for the energy transfer is not dependent on $|\bm{\kappa}|$. These results agree with those reported in the literature, where \citet{mininniLargescaleFlowEffects2006} showed a collapse of the transfer functions around the $|\bm{\kappa}| - K'$ shell.

The distribution of \mtm energy transfers is consistent with previously obtained results and clarifies that the scale-local energy transfer in nonlocal interactions are the most intense on individual basis. We now proceed to link these results to the potential $\hat{Q}$.

\subsection{\textit{Catalytic} vs. \textit{reacting} energy transfers}
\label{sec:LWFCatalyticVSReacting}

\begin{figure}[t]
    \centering
    \subfloat[{$\bm{\kappa} = [30,0]$}\label{subfig:LWF_M2M_QT_30}]{%
        \includegraphics[]{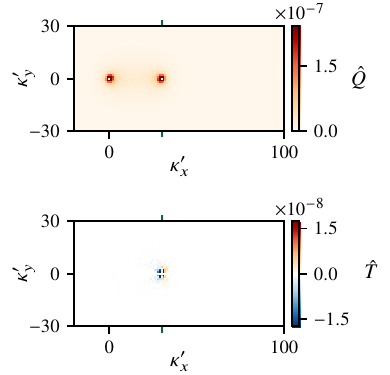}
    }
    \hspace{2cm}
    \subfloat[{$\bm{\kappa} = [60,0]$}\label{subfig:LWF_M2M_QT_60}]{%
        \includegraphics[]{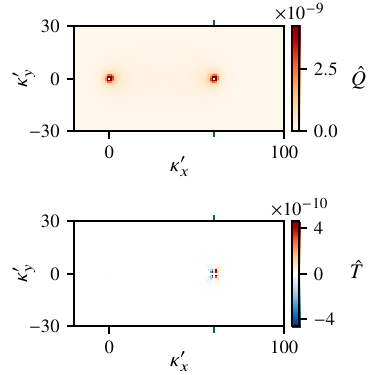}
    }
    \caption{Comparison of potential $\hat{Q}$ (top rows) and \mtm energy transfer $\hat{T}$ (bottom rows) for the LWF case ($\kappa_{f} \sim 1$). The green dashes indicate the position of the sampling point.}\label{fig:LWF_M2M_QT}
\end{figure}

Upon the introduction of the potential $\hat{Q}$, we mentioned that it is symmetric in its arguments. Indeed, for a given triad with fixed $\kP$ and $\kmkP$, we have $\hat{Q}(\k, \kP) = \hat{Q}(\k, \kmkP)$ since it is only based on the energy content of the modes involved. When considering in particular a nonlocal triad, it means that the potential of scale local (between $\k$ and $\kP$) and scale nonlocal (between $\k$ and $\kmkP$) energy transfer is the same. This is illustrated in the top rows of Fig. \ref{fig:LWF_M2M_QT}, which shows the regions of high potential, one centered around the origin and the other centered around the sampling point.

However, the direct comparison of the potential $\hat{Q}$ to the \mtm energy transfer $\hat{T}$ reveals that the HET kernel is only composed of the \textit{catalytic} energy transfers, as described in Sec. \ref{sec:ComparisonLiterature}, while the \textit{reacting} energy transfers are much smaller. We recover here that the most intense energy transfers occur in a nonlocal interaction, but that within this interaction only the local energy transfer is large and not the nonlocal one. Since our proposal is that the most intense energy transfers are the ones with the largest potential and that both local and nonlocal energy transfer share the same potential, we provide arguments to explain the dampening of the \textit{reacting} energy transfers.

\begin{figure}[t]
    \centering
    \subfloat[{$\bm{\kappa} = [30,0]$}\label{subfig:LWF_eta_30}]{%
        \includegraphics[]{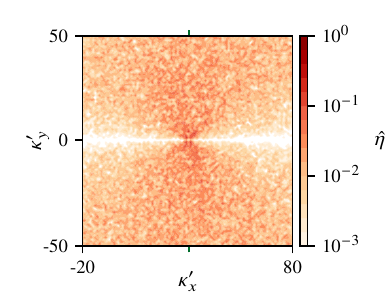}
    }
    \hspace{2cm}
    \subfloat[{$\bm{\kappa} = [60,0]$}\label{subfig:LWF_eta_60}]{%
        \includegraphics[]{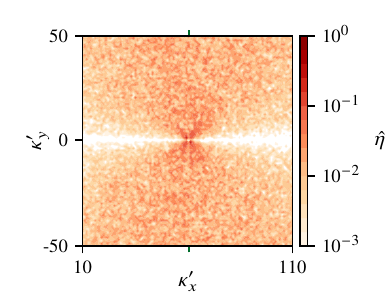}
    }
    \caption{Energy transfer effectiveness for the LWF case ($\kappa_{f} \sim 1$) centered around the sampling point. The green dashes indicate the position of the sampling point.}\label{fig:LWF_eta}
\end{figure}

For this purpose, we introduce the effectiveness of the \mtm energy transfer as the ratio of actual (averaged) transfer $\hat{T}$ to the potential $\hat{Q}$ as
\begin{equation}
    \hat{\eta}(\k, \kP) = \frac{|\hat{T}(\k, \kP)|}{\hat{Q}(\k, \kP)}.
    \label{eq:effectiveness}
\end{equation}
Figure \ref{fig:LWF_eta} shows the effectiveness of energy transfers of the sampling points presented in Fig. \ref{fig:LWF_M2M_QT}. The first observation is that the patterns are similar for both sampling points. In fact, these patterns are similar for all sampling points (not shown here), and while we discuss the dependence of distribution of $\hat{T}$ with different forcing configurations in the following sections, $\hat{\eta}$ remains the same in these cases, see App. \ref{app:Effectiveness}. This leads us to believe the effectiveness patterns show some universality, independent of the shape of the energy spectrum.

\begin{figure}[t]
    \centering
    \subfloat[\label{subfig:dampenedTriadInteraction}]{%
        \includegraphics[]{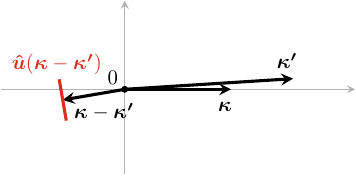}
    }
    \hspace{2cm}
    \subfloat[\label{subfig:favouredTriadInteraction}]{%
        \includegraphics[]{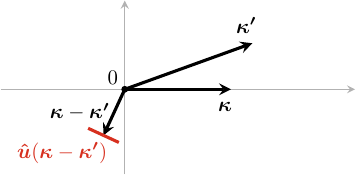}
    }
    \caption{Dampened (a) and favoured (b) triad interactions as a result of the divergence-free condition.}\label{fig:favouredTriadInteraction}
\end{figure}

We identify ``cones of low effectiveness'' regions along the $\kappa'_{x}$ axis on each side of the sampling point (conical in three dimensions due to axial symmetry) and associate them to effects of the divergence-free condition (Eq. \ref{eq:DivFree}). In the expression for the \mtm energy transfer $\hat{T}$ (Eq. \ref{eq:ModeToModeEnergyTransfer}) appears the factor $\kappa_{l} \hat{u}_{l}(\bm{\kappa} - \bm{\kappa'})$, which is zero if $\k$ is normal to $\bm{\hat{u}}(\bm{\kappa} - \bm{\kappa'})$. The divergence-free condition states that the velocity mode is normal to its wavenumber, such that $\bm{\hat{u}}(\bm{\kappa} - \bm{\kappa'})$ is normal to $\kmkP$. Therefore, if $\k$ and $\kP$ are such that $\k$ is close to aligned with $\kmkP$ (i.e., $\kP$ is near the $\kappa'_{x}$ axis), the result is that $\k$ is also close to normal to $\bm{\hat{u}}(\bm{\kappa} - \bm{\kappa'})$, thus effectively dampening the energy transfer, as sketched in Fig. \ref{fig:favouredTriadInteraction}. This behaviour is encompassed in the EDQNM estimator through the geometric factor $b$. Figure \ref{subfig:ModelSpectrumLWFEDQNMbSlice} shows low values of $b$ along the $\kappa'_{x}$ axis and values close to 1 on the sides ($\k$ and $\kP$ share the same $\kappa'_{x}$ component).

Furthermore, there is a decay of $\hat{\eta}$ when $|\kmkP|$ is growing. In the EDQNM estimator, it is the $\theta(\bm{\kappa}, \bm{\kappa'})$ function that plays this role. The eddy-damping (or eddy relaxation) parameter is modeled as such since ``the memory should cut off because phase correlations between Fourier modes (or rather eddies) do not persist in the random convection fields of other eddies. Coherence is destroyed by non-linear scrambling'' \citep{orszagAnalyticalTheoriesTurbulence1970}. In short, neighbouring modes are expected to be phase correlated such that $|\hat{T}|$ is maximized, and this correlation weakens as the modes are further away. Recent results \citep{kangAlignmentsTriadPhases2021,protasAlignmentsTriadPhases2024} show that ``triads involving energy-containing Fourier modes align their phases so as to maximize the energy flux toward small scales'' in the one dimensional Burgers equation, namely that the interactions with neighbouring modes display high level of phase correlation. It can be expected that this result also holds for the three dimensional Navier-Stokes equations, which will be the subject of subsequent studies.

These two effects combine such that, while having a large potential, the \textit{reacting}  energy transfers are small (although not always negligible, see Sec. \ref{sec:ResidualNonlocalEnergyTransfers}), as expected by the K41 theory. Indeed, the ECR is usually around the origin (large scales), and therefore for any direction of $\k$, the corresponding $\kmkP$ mode will always be in the ``cone of low effectiveness'', where $\k$ is normal to $\bm{\hat{u}}(\bm{\kappa} - \bm{\kappa'})$.

Having now established that the \mtm energy transfer $\hat{T}$ is dependent on the potential $\hat{Q}$, modulated by dampening factors, we note that the potential $\hat{Q}$ is itself determined by the energy spectrum. This naturally leads to the question of the dependency of $\hat{T}$ on the energy spectrum.

\section{Influence of large scales on the extent of energy transfer locality}
\label{sec:ExtentOfLocality}
In this section, we analyse the distribution of the \mtm energy transfer $\hat{T}$ in cases with shifted ECR, produced by forcing in a specific range; see Tab. \ref{tab:SimulationParameters}.

\begin{figure}[t]
    \centering
    \subfloat[{}\label{subfig:ModelSpectrumIWFEnergySpectrum}]{%
        \includegraphics[]{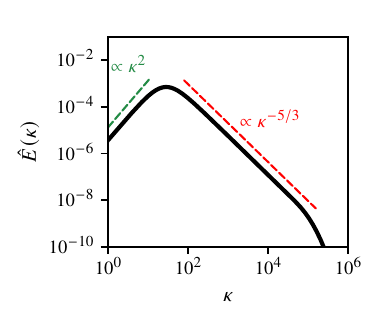}
    }
    \hspace{2cm}
    \subfloat[{}\label{subfig:ModelSpectrumIWFPotentialLinear}]{%
        \includegraphics[]{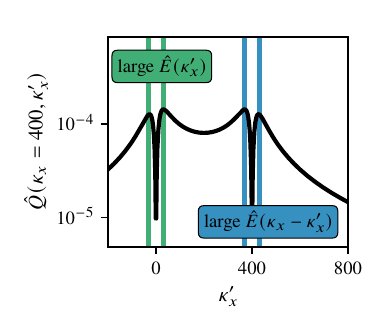}
    }
    \caption{Model kinetic spectrum with energy at intermediate scales (a) and resulting \mtm potential $\hat{Q}$ (b).}\label{fig:ModelSpectrumIWFScaling}
\end{figure}

\subsection{Potential with shifted energy containing range}

When the forcing is not applied directly at the largest available scales in the domain, the energy spectrum presents a thermal equilibrium range, with a scaling proportional to $\kappa^{2}$. Figure \ref{fig:ModelSpectrumIWFScaling} presents such a spectrum computed again with the model presented in \citet{pope2000turbulent}, determined by the energy containing scale $L = L_{0}/20$, the viscosity $\nu = 10^{-7}$ and $\Rey_{\lambda} = 10^{5}$. The right panel shows the resulting potential along $\kappa'_{x}$. Since the energy containing range is now a shell in spectral space, the regions with high potential (both \textit{reacting} and \textit{catalytic}) are now also shells, centered around the origin and the sampling point, respectively.

This means that the \mtm energy transfer, if indeed strongly determined by the potential as proposed in the previous section, will not be the most intense with direct neighbours, but instead with modes slightly further away. The HET kernel will be similar to the imprint of the ECR shifted around the sampling point.

\subsection{Distribution of spectral energy transfers}
\label{sec:IWFDistributionOfT}

\begin{figure}[t]
    \centering
    \subfloat[{$\bm{\kappa} = [30,0]$, $\kappa_{f} \sim 10$}\label{subfig:IWF_10_M2M_QHT_30}]{%
        \includegraphics[]{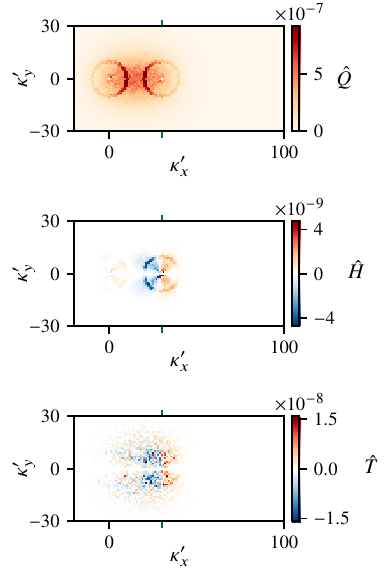}
    }
    \hspace{2cm}
    \subfloat[{$\bm{\kappa} = [60,0]$, $\kappa_{f} \sim 10$}\label{subfig:IWF_10_M2M_QHT_60}]{%
        \includegraphics[]{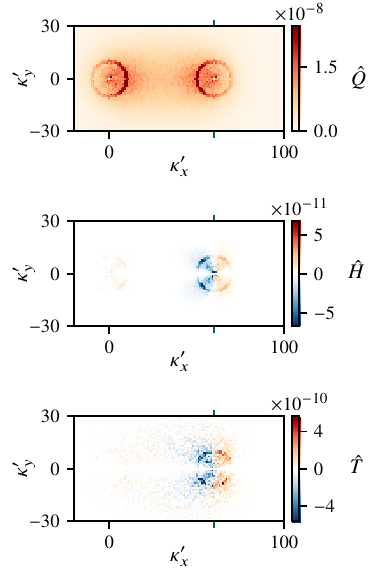}
    }
    \caption{Comparison of potential $\hat{Q}$ (top rows), EDQNM estimator $\hat{H}$ (middle rows) and \mtm energy transfer $\hat{T}$ (bottom rows) for the IWF10 case ($\kappa_{f} \sim 10$). The green dashes indicate the position of the sampling point.}\label{fig:IWF10_M2M}
\end{figure}

\begin{figure}[t]
    \centering
    \subfloat[{$\bm{\kappa} = [60,0]$, $\kappa_{f} \sim 20$}\label{subfig:IWF20_M2M_QHT_60}]{%
        \includegraphics[]{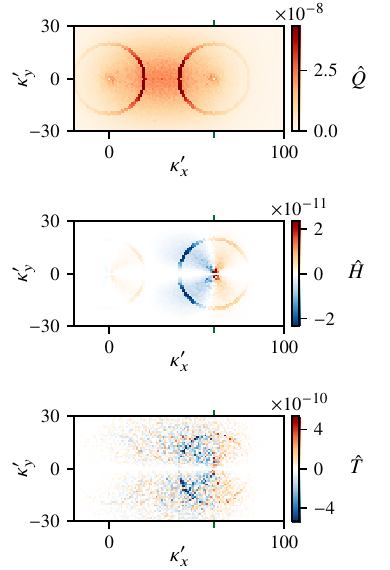}
    }
    \hspace{2cm}
    \subfloat[{$\bm{\kappa} = [80,0]$, $\kappa_{f} \sim 20$}\label{subfig:IWF20_M2M_QHT_80}]{%
        \includegraphics[]{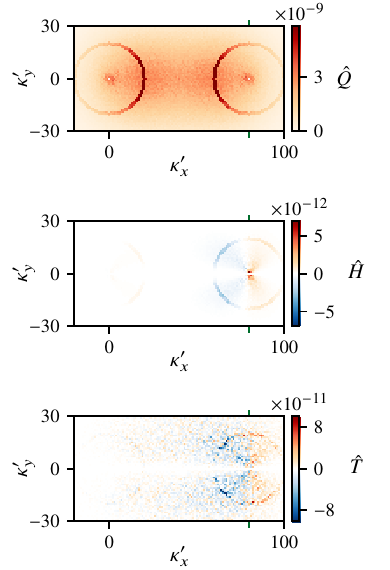}
    }
    \caption{Comparison of potential $\hat{Q}$ (top rows), EDQNM estimator $\hat{H}$ (middle rows) and \mtm energy transfer $\hat{T}$ (bottom rows) for the IWF20 case ($\kappa_{f} \sim 20$). The green dashes indicate the position of the sampling point.}\label{fig:IWF20_M2M}
\end{figure}

Figures \ref{fig:IWF10_M2M} and \ref{fig:IWF20_M2M} present the distributions of the potential $\hat{Q}$, the EDQNM estimator $\hat{H}$ and of the \mtm energy transfer $\hat{T}$ for the cases IWF10 ($\kappa_{f} \sim 10$) and IWF20 ($\kappa_{f} \sim 20$), respectively. The sampling points are chosen such that the \textit{reacting} and \textit{catalytic} regions (as seen in the potential $\hat{Q}$ figures) are distinct. The overlapping cases are discussed in Sec. \ref{sec:ResidualNonlocalEnergyTransfers}.

The main observation is that the HET kernel does indeed match the \textit{catalytic} region. The most intense transfers correspond to the high potential region with the \textit{catalyst} mode in the ECR, as indicated by the rings of radii 10 and 20, respectively, for the IWF10 and IWF20 cases, observed in the \mtm energy transfer panels.

This confirms the proposition that $\hat{T}$ is strongly dependent on the spectral location of the ECR, and allows us to revise the classical argument that the most intense interactions correspond to ``local energy transfer in nonlocal interactions''. The intensity of the energy transfer is not related to whether the interaction is local or nonlocal, but rather to the energy content of the modes involved. In most simulations, the ECR is at the largest available scales and therefore the intense energy transfers also correspond to nonlocal interactions as described in the literature \citep{domaradzkiLocalEnergyTransfer1990,waleffeNatureTriadInteractions1992}.

\begin{figure}[t!]
    \centering
    \includegraphics[]{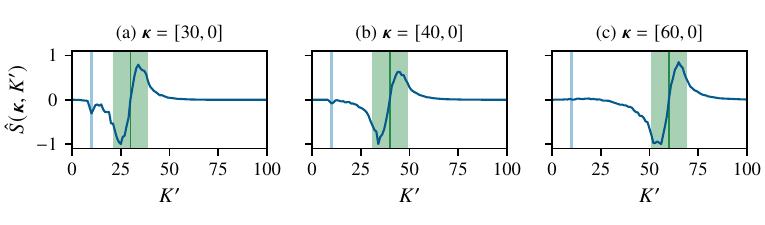}
    \caption{\textit{Shell-to-shell} transfer functions $\hat{S}(\kappa, K')$ for the IWF10 case. The blue shaded area correspond to the ECR, the green shaded area to the range $]\kappa_{x} - \kappa_{f}, \kappa_{x} + \kappa_{f}]$ and the vertical green line to the sampling point.}\label{fig:IWF10_M2S}
\end{figure}

\begin{figure}[t!]
    \centering
    \includegraphics[]{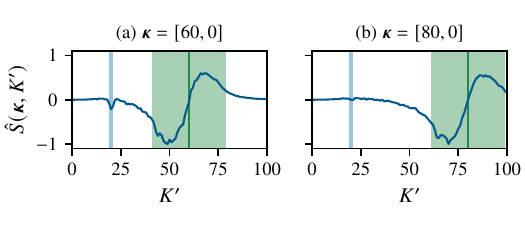}
    \caption{\textit{Shell-to-shell} transfer functions $\hat{S}(\kappa, K')$ for the IWF20 case. The blue shaded area correspond to the ECR, the green shaded area to the range $]\kappa_{x} - \kappa_{f}, \kappa_{x} + \kappa_{f}]$ and the vertical green line to the sampling point.}\label{fig:IWF20_M2S}
\end{figure}

The comparison between the \mtm energy transfer $\hat{T}$ and the EDQNM estimator $\hat{H}$ shows reasonable agreement, both in distribution and intensity. However, and specifically in the IWF20 case, it seems that $\hat{H}$ is slightly overestimating the intensity of energy transfer to neighbouring modes, which may even surpass the transfer in the \textit{catalytic} region; see (Fig. \ref{subfig:IWF20_M2M_QHT_80}). As explained in Sec. \ref{sec:LWFCatalyticVSReacting}, the energy transfer of neighbours is enhanced through the $\theta$ function (Eq. \ref{eq:EDQNMtheta}). The choice of model for the eddy-damping parameter is thus probably the cause of this discrepancy.

It appears clearly that the energy transfer displays the same specular behaviour predicted by the EDQNM estimator and discussed for the LWF case in Sec. \ref{sec:LWFCatalyticVSReacting}. Indeed, the energy transfers are suppressed along the $\kappa'_{x}$ axis, and the analysis of the effectiveness patterns in App. \ref{app:Effectiveness} reveals that the ``cones of low effectiveness'' are also observed in both IWF10 and IWF20 cases. The external cone (for $\kappa'_{x} > \kappa_{x}$) appears to be wider than the inner one (for $\kappa'_{x} < \kappa_{x}$). This is attributed to the observation that the potential $\hat{Q}$ is larger towards the larger scales than the smaller scales, and less to the suppression patterns of the divergence-free condition being different on either side of $\kappa_{x}$. This behaviour can also be seen in Figs. \ref{subfig:ModelSpectrumLWFPotentialLinear} and \ref{subfig:ModelSpectrumIWFPotentialLinear}, where the regions between the peaks of $\hat{Q}$ show smaller decay than the outer ones.

Finally, the forward nature of the convection term is established for all sampling points, where the energy transfers are negative for $|\kP| < |\k|$ and positive for $|\kP| > |\k|$. It is visible on the EDQNM estimator $\hat{H}$ panels that the energy transfers are close to zero for $|\kP| \approx |\k|$. In these regions the energy transfers are also intense but not sign definite, and therefore average to zero contribution. This result is expected to hold for the \mtm energy transfer $\hat{T}$ upon sufficient convergence of the statistics.

Figures \ref{fig:IWF10_M2S} and \ref{fig:IWF20_M2S} show the \sts transfer functions for the IWF10 and IWF20 cases, respectively. As can be expected from the distribution of energy transfers presented above, the transfer functions are spread over several shells, with most of the transfers happening with shells within the range $]\kappa_{x} - \kappa_{f}, \kappa_{x} + \kappa_{f}]$ (indicated in light green).

The locality of energy transfer holds as the exchanges are primarily with neighbouring shells. The extent of locality, which can be defined as the distance between the sampling mode and the peaks of the transfer functions, scales with $\kappa_{f}$. Note that this extent of locality is not the same on either side of the sampling mode, which is again linked to the bias of the potential $\hat{Q}$ towards the large scales. Furthermore, the peaks of $\hat{S}(\kappa, K')$ are not exactly at $\kappa_{x} \pm \kappa_{f}$. This was already observed in \citet{mininniLargescaleFlowEffects2006} and attributed to moderate Reynolds number effects. Here we argue that it is due to the fact that the \sts is obtained from progressive shell integration through the \textit{catalytic} region. We have established the specular behaviour of the latter, and it is therefore to be expected that the extent of locality scales with $\kappa_{f}$, but is not directly proportional to it.

\subsection{Residual nonlocal energy transfers}
\label{sec:ResidualNonlocalEnergyTransfers}

We have seen in the previous sections that the \textit{reacting} energy transfers (direct exchange with the large scales), while having a large potential, are dampened out by geometric and phase-correlation factors. However, the suppressing factors may not always be sufficient and residual nonlocal energy transfers are observed. This behaviour is barely visible on the $\hat{H}$ and $\hat{T}$ panels in Figs. \ref{fig:IWF10_M2M} and \ref{fig:IWF20_M2M}, with non-zero energy transfers with the ECR, but clearly appears in the \sts transfer function aggregates $\hat{S}$, Figs. \ref{fig:IWF10_M2S} and \ref{fig:IWF20_M2S}. Indeed, there is non-zero energy transfer with the ECR shell (here corresponding to the forcing range, indicated in blue). Note that this transfer is negative, such that on average energy is transferred from the large scales directly to the sampling point.

\begin{figure}[t]
    \centering
    \subfloat[{$\bm{\kappa} = [30,0]$, $\kappa_{f} \sim 20$}\label{subfig:IWF20_M2M_HT_M2S_30}]{%
        \includegraphics[]{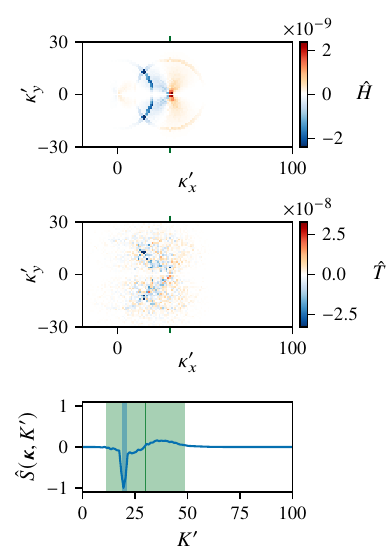}
    }
    \hspace{2cm}
    \subfloat[{$\bm{\kappa} = [40,0]$, $\kappa_{f} \sim 20$}\label{subfig:IWF20_M2M_HT_M2S_40}]{%
        \includegraphics[]{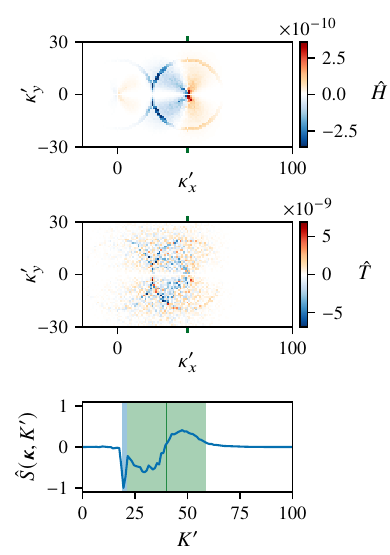}
    }
    \caption{Comparison of EDQNM estimator $\hat{H}$ (top rows), \mtm energy transfer $\hat{T}$ (middle rows) and resulting \sts transfer function $\hat{S}$ (bottom rows) for the IWF20 case ($\kappa_{f} \sim 20$). The green dashes indicate the position of the sampling point.}\label{fig:IWF20_M2M_HT_M2S}
\end{figure}

If the sampling mode is close enough to the ECR, it reveals direct interplay between the \textit{reacting} and \textit{catalytic} interactions, as shown in Fig. \ref{fig:IWF20_M2M_HT_M2S}. On the left panel the two regions of high potential intersect, resulting in dominant direct energy transfer with the large scales. These residual transfers are described as nonlocal because they represent energy transfer to fixed modes (in the ECR) independently of the sampling mode location, whereas the \textit{catalytic} region is always shifted around the sampling point. One may still consider them as local in the sense that the modes of the ECR are within the range $]\kappa_{x} - \kappa_{f}, \kappa_{x} + \kappa_{f}]$. Progressively, as $|\k|$ increases, the amount of nonlocal direct energy exchanges with the ECR decreases to zero.

Such direct energy transfers with the ECR can also be seen in the literature, e.g., in Fig. 14 of \citet{mininniLargescaleFlowEffects2006}. As is the case here, their intensity decreases against growing $Q$ [here $K'$].

\subsection{Summary of the distribution of energy transfers}

\begin{figure}[t]
    \centering
    \includegraphics[]{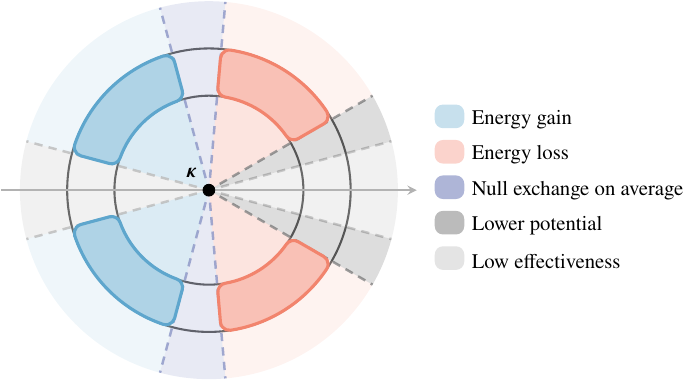}
    \caption{Sketch of the spectral distribution of the \mtm energy transfer $\hat{T}$ with the ECR at some intermediate scales.}\label{fig:summaryDistributionT}
\end{figure}

Figure \ref{fig:summaryDistributionT} presents a sketch of the distribution of the \mtm energy transfer $\hat{T}(\k, \kP)$, assuming the residual nonlocal energy transfers presented in the previous section are negligible. Note that in three dimensions the resulting distribution is axisymmetric around the axis given by the direction of the sampling point. The most intense energy transfers are located in the \textit{catalytic} interactions region, which is an imprint of the ECR shifted around the sampling point.

The specular behaviour is explained as follows: for $|\kP| < |\k|$ the transfers are on average negative, for $|\kP| \approx |\k|$ the transfers are not sign definite and average to zero and for $|\kP| > |\k|$ the transfers are on average positive. Furthermore, the transfers are suppressed along the axis given by the direction of $\k$ due to geometric factors and a bias of the potential $\hat{Q}$ towards the larger scales.

\subsection{Open questions}

What is established in this work is that the most intense \mtm energy transfers are strongly dependent on the spectral location of the most energetic scales. In most cases, the latter are equivalent to the largest scales considered, such that the most intense transfers are associated to interactions described as nonlocal in the literature \citep{domaradzkiLocalEnergyTransfer1990,waleffeNatureTriadInteractions1992}. However, it does not mean that these interactions will dominate the total energy transfer and thus the dynamics of the turbulent cascade. Indeed, as objected in \citet{zhouDegreesLocalityEnergy1993,zhouInteractingScalesEnergy1993} and \citet{eyinkEnergyDissipationViscosity1994,eyinkLocalEnergyFlux1995}, the number of these intense energy transfers is fixed and not dependent on the sampling point wavenumber $\kappa$. If we define a local interaction in the spirit of \citet{kraichnanInertialrangeTransferTwo1971}, that is, such that the minimum wavenumber in the triad is more than half the sampling point wavenumber $\kappa$, and the maximum wavenumber in the triad is less than twice $\kappa$, then the set of local interactions $\mathcal{L}_{\mathrm{local}}$ is given by the intersection of two spectral domains as
\begin{equation}
    \mathcal{L}_{\mathrm{local}}(\k) = \left\{ \kP \mid 0.5 \kappa < |\kP| < 2 \kappa \right\} \cap \left\{ \kP \mid 0.5 \kappa < |\kmkP| < 2 \kappa \right\}
\end{equation}
As both domains are spherical shells and their intersection is guaranteed by construction, the total number of local interactions grows like $\sim \kappa^{3}$. If the intensity of the \mtm energy transfers does not decay too fast against growing distance $|\kmkP|$, then the local interactions will eventually contribute more to the total energy transfer than the intense nonlocal interactions. Determining a scaling of $\hat{T}(\k, \kP)$ on $|\kmkP|$ from the data is difficult due to the specular behaviour discussed in the previous sections. The next step is then to compare the contribution of the HET kernel and the local interaction region to the total energy transfer against $\k$ for different Reynolds number, for which a new set of simulations with wider inertial range is necessary.

The decay of residual nonlocal interactions can be studied in a similar manner. While the present work shows that these residuals become small already at moderate Reynolds numbers, they may explain why earlier results showed how small scales are tied to the large scale dynamics \citep{yeungResponseIsotropicTurbulence1991,biferaleIsotropyVsAnisotropy2001,biferaleAnisotropicHomogeneousTurbulence2001}.

The computation of individual \mtm energy transfer also opens the possibility to study in depth the phase correlation between modes. Recent results have shown that neighbouring modes are phase correlated such that the energy transfer is maximized towards the smaller scales with viscous Burgers equation. It remains to be established if this ``phase correlation locality'' also holds in three dimensional Navier-Stokes turbulence, and whether it is associated to the \textit{catalytic} region presented in this article. If the latter is verified, it is also interesting to check if the neighbouring modes which are phase correlated to each other are also correlated to the \textit{catalyst} mode.

The study of phase correlation is still highly relevant. Indeed, as discussed in \citet{orszagAnalyticalTheoriesTurbulence1970,leithAtmosphericPredictabilityTwoDimensional1971,lesieurTurbulenceFluids2008}, the only modelling part in the EDQNM approach (apart from the quasi-normal approximation which is justified therein) is the eddy damping parameter in the function $\theta(\k, \kP)$ (Eq. \ref{eq:EDQNMtheta}), which is artificially added to mimic the destruction of coherence (phase correlation) by the nonlinear convection. Better understanding of the distribution and scaling of the phase correlation can thus lead to improved models for spectral viscosity, which are still relevant today for Large Eddy Simulation subgrid scale modelling.

\section{Conclusions}

We present direct computation of individual \mtm energy transfers $\hat{T}(\bm{\kappa}, \bm{\kappa}')$ from DNS of homogeneous isotropic turbulence. This approach differs from previous investigations, which relied on shell-filtered transfer functions that could not distinguish between the energy exchanged by individual modes within a triad. Our methodology allows for a clear identification of the \textit{reacting} mode (which exchanges energy with the primary mode) and the \textit{catalyst} mode (which mediates the interaction without direct energy exchange). The comparison between $\hat{T}$ and an EDQNM function $\hat{H}$ shows good agreement in all cases, recovering the forward, scale-local nature of energy transfer consistent with the cascade picture.

We introduce the potential $\hat{Q}(\bm{\kappa}, \bm{\kappa}')$, a dimensional estimate of the maximum energy transfer based on the energy content of the modes involved. The fact that $\hat{Q}$ successfully predicts the distribution of $\hat{T}$ in the region around the sampling point allows us to link the intensity of energy transfers to the spectral location of the energy containing range. This provides a new perspective on the classical finding of ``local energy transfer in nonlocal interactions'': such interactions dominate not because they are nonlocal, but because they involve a catalyst mode in the energy containing range.

The potential also predicts direct, nonlocal, exchanges with modes in the energy containing range. Our analysis shows that these \textit{reacting} transfers are suppressed by geometric factors arising from the divergence-free condition, with phase correlation likely playing an additional damping role, as is the case in the EDQNM theory. Despite this suppression, residual nonlocal energy transfers persist when the sampling mode is close to the energy containing range.

To further test our proposal, we perform DNS with forcing at intermediate wavenumbers. In these cases the region of intense energy transfers shifts accordingly, forming a ring (sphere in three dimensions) around the sampling point whose radius scales with the forcing wavenumber $\kappa_f$. The scale locality of the energy transfer is also recovered upon shell integration of the individual contributions. The analysis of the similar patterns displayed between the \mtm energy transfer $\hat{T}$, the potential $\hat{Q}$ and the EDQNM function $\hat{H}$ in all presented cases validates the proposal that the energy content of the modes involved in the triad is crucial to the intensity of the energy transfer quantities.

While the results confirm that what is known as nonlocal triad interactions do dominate in intensity, the natural next step is to use a similar approach to recover the scaling of the contribution of such interactions to total energy transfer quantities, which is expected to decrease as the scales get smaller, according to the central assumption of universality of small scales in turbulent flows.

\clearpage
\appendix

\section{Kinetic energy spectra}
\label{app:kineticEnergySpectra}

\begin{figure}[h!]
    \centering
    \subfloat[{$\kappa_{f} \sim 1$}\label{subfig:LWF_kinenticEnergySpectrum}]{%
        \includegraphics[]{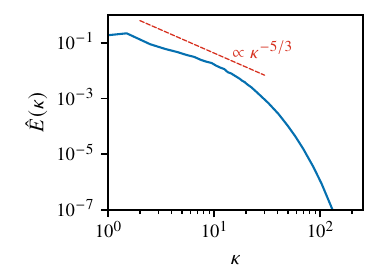}
    }
    \hspace{2cm}
    \subfloat[{$\kappa_{f} \sim 10$}\label{subfig:IWF10_kinenticEnergySpectrum}]{%
        \includegraphics[]{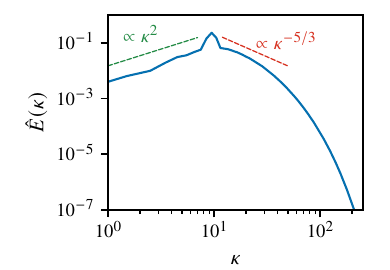}
    }
    \hspace{2cm}
    \subfloat[{$\kappa_{f} \sim 20$}\label{subfig:IWF20_kinenticEnergySpectrum}]{%
        \includegraphics[]{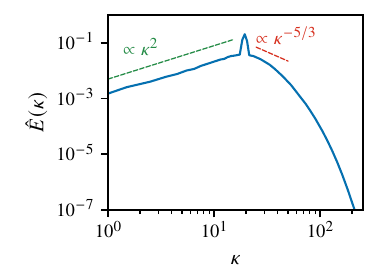}
    }
    \caption{Energy spectra for the LWF (a), IWF10 (b) and IWF20 (c) cases.}\label{fig:kineticEnergySpectra}
\end{figure}

Figure. \ref{fig:kineticEnergySpectra} presents the energy spectra of the cases studied.

\section{Effectiveness of \mtm energy transfers}
\label{app:Effectiveness}

\begin{figure}[t]
    \centering
    \subfloat[{$\bm{\kappa} = [30,0]$, $\kappa_{f} \sim 10$}\label{subfig:IWF10_eta_30}]{%
        \includegraphics[]{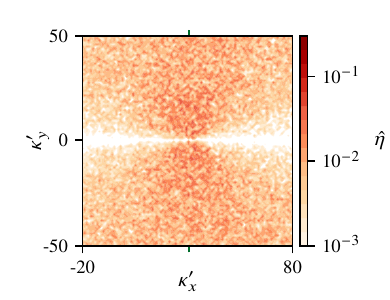}
    }
    \hspace{2cm}
    \subfloat[{$\bm{\kappa} = [60,0]$, $\kappa_{f} \sim 10$}\label{subfig:IWF10_eta_60}]{%
        \includegraphics[]{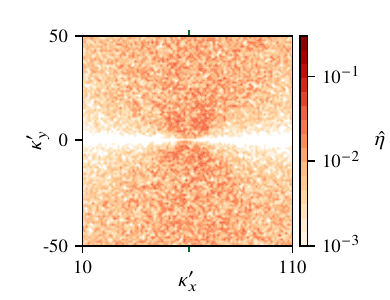}
    }
    \\
    \subfloat[{$\bm{\kappa} = [30,0]$,$\kappa_{f} \sim 20$}\label{subfig:IWF20_eta_30}]{%
        \includegraphics[]{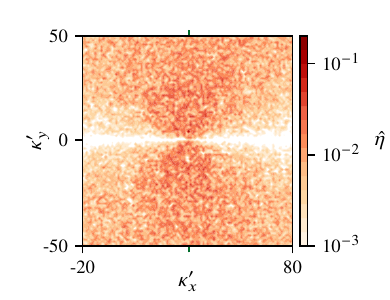}
    }
    \hspace{2cm}
    \subfloat[{$\bm{\kappa} = [60,0]$, $\kappa_{f} \sim 20$}\label{subfig:IWF20_eta_60}]{%
        \includegraphics[]{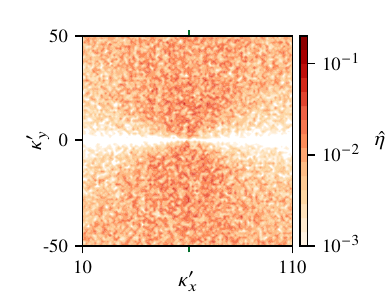}
    }
    \caption{Energy transfer effectiveness for the IWF10 case ($\kappa_{f} \sim 10$, top row) and IWF20 case ($\kappa_{f} \sim 20$, bottom row) centered on the sampling point. The green dashes indicate the position of the sampling point.}\label{fig:IWF_eta}
\end{figure}

Figure. \ref{fig:IWF_eta} presents the energy transfer effectiveness $\hat{\eta}$ (Eq. \ref{eq:effectiveness}) for the IWF10 and the IWF20 cases, on the top and bottom rows respectively.

\section{Efficient computation of spectral energy transfer $\hat{T}$}\label{app:EnergyTransferComputation}

Given a sample point $\bm{\kappa}^{(s)}$, the \textit{mode-to-mode} energy transfer $\hat{T}(\bm{\kappa}^{(s)}, \bm{\kappa'})$ is computed directly as in Eq. \ref{eq:ModeToModeEnergyTransfer}. The key step involves obtaining the shifted velocity field $\hat{\bm{u}}(\bm{\kappa}^{(s)} - \bm{\kappa'})$ from the solution variable $\hat{\bm{u}}(\bm{\kappa'})$. In the naive implementation, one can simply loop over all $\bm{\kappa'}$ in the domain and compute the velocity field shifted at $\bm{\kappa}^{(s)} - \bm{\kappa'}$. This is highly inefficient and renders the computation unfeasible for large spectral domains. Instead, using the fact that the velocity field is also periodic in spectral space (that is, aliased due to the use of FFT operations), the whole shifted velocity field can be obtained at once using \texttt{roll} and \texttt{flip} operations, as described in the following algorithm (making use of \texttt{Python} negative indexing)\\

\begin{enumerate}
    \item Compute $\hat{\bm{u}}(-\bm{\kappa'})$ by flipping the original velocity field $\hat{\bm{u}}(\bm{\kappa'})$. Note that there is also a \texttt{roll} operation of one index in every direction, which is required to keep the index $[0,0,0]$ at the origin of the spectral domain in case of even resolution $N$\\

          \hspace{1.5cm}$\hat{\bm{u}}(-\bm{\kappa'}) \leftarrow \texttt{flip} \bigl[\ \texttt{roll}[\hat{\bm{u}}(\bm{\kappa'}), \ \texttt{shifts} = (-1, -1, -1)]\ \bigr]$

          \hspace{0.2cm}\item Compute $\hat{\bm{u}}(\bm{\kappa}^{(s)} - \bm{\kappa'})$ by shifting the velocity field $\hat{\bm{u}}(-\bm{\kappa'})$ obtained in the previous step according to the chosen sampling point\\

          \hspace{1.5cm}$\hat{\bm{u}}(\bm{\kappa}^{(s)} - \bm{\kappa'}) \leftarrow \texttt{roll}[\hat{\bm{u}}(- \bm{\kappa'}), \ \texttt{shifts} = \bm{\kappa}^{(s)}]$

          \hspace{0.2cm}\item Dealias $\hat{\bm{u}}(\bm{\kappa}^{(s)} - \bm{\kappa'})$ as the previous shift operation introduce non-physical periodicity in the spectral velocity field. This is done by setting to zero all modes with wavenumbers $\bm{\kappa}^{(s)} - \bm{\kappa'}$ outside the spectral domain.
\end{enumerate}\vspace{0.5cm}

The final \textit{mode-to-mode} energy transfer $\hat{T}(\bm{\kappa}^{(s)}, \bm{\kappa'})$ is then computed as a matrix multiplication of the shifted velocity field $\hat{\bm{u}}(\bm{\kappa}^{(s)} - \bm{\kappa'})$ and the original velocity field $\hat{\bm{u}}(\bm{\kappa'})$ (plus summation for the dot product), significantly reducing the computational cost compared to the naive implementation. Note that the same approach is used to obtain the shifted energy field $\hat{E}(\bm{\kappa}^{(s)} - \bm{\kappa'})$ for the energy transfer potential (Eq. \ref{eq:ModeToModeEnergyTransferPotential}) and the EDQNM function $\hat{H}$ (Eq. \ref{eq:EDQNMH}).

\bibliographystyle{plainnat}
\bibliography{references.bib}

\end{document}